\documentclass[%
 reprint,
 amsmath,amssymb,
 aps,
]{revtex4-1}

\usepackage{graphicx}
\usepackage{dcolumn}
\usepackage{bm}

\begin{document}


\title{Patterns of Gendered Performance Difference in Introductory STEM Courses}

\author{Benjamin P. Koester}
 \email{bkoester@umich.edu}
\author{Galina Grom}
 \email{grom@umich.edu}
\author{Timothy A. McKay}
\email{tamckay@umich.edu}
    \affiliation{Department of Physics, University of Michigan, 450 Church St., Ann Arbor, MI 48109, USA}%

\date{\today}

\begin{abstract}
Grades provide students with their primary performance feedback: signals which affect academic choices.  Variations in grading practice among courses impose grade penalties (and bonuses) on students who take them. These grade penalties are sometimes gendered. Using extensive data from the University of Michigan, we report on patterns of grade penalty and gendered performance difference across 116 large courses. We find that significant gendered performance differences are ubiquitous in large introductory STEM lecture courses. They are largely absent in both STEM labs and in lecture courses in other disciplines. Exploring the features of these courses, we hypothesize that evaluation methods used in STEM lecture courses interact with stereotype threat to create these gendered performance differences.
\end{abstract}

\maketitle


\section{\label{sec:level1}Introduction}

Despite generations of gradual progress, women remain underrepresented in the leadership of all STEM disciplines \cite{committee2010gender}. The causes of these disparities in participation are certainly various, but one is the existence of gendered performance differences (GPDs) in introductory STEM courses \cite{Ost2010-qj,kost2009characterizing,kost2010gender,lorenzo2006reducing,pollock2007reducing,kost2011characterizing}. Significant GPDs exacerbate the already low average grades students receive in these courses, sending signals of incompetence and challenging efforts to diversify science. 
Identity based performance differences are common in STEM. They raise important questions of equity in higher education.

Methods used to measure performance differences vary. Some simply examine outcomes, comparing grades or scores on concept inventories for male and female students at one point in time. Others make efforts to account for background and preparation, either through pre-and-post testing, or by including measures of prepation in performance models (e.g. high school GPA, standardized tests, and prior college performance) \cite{riegle2012more}. Most studies of performance gaps have been narrow in scope, exploring single courses in individual subjects for limited periods of time. Comparison among these studies is complicated by their various contexts and analytic approaches.

These limitations leave many important questions unaddressed. Is the origin of GPDs most closely related to subject matter, instructional style, student background, or mode of assessment? To obtain initial insight into these many possibilities, we have conducted uniform measurements of gendered performance difference across a multi-disciplinary array of 116 large enrollment courses at the University of Michigan. These courses vary substantially in subject matter, instructional style, and mode of assessment. As at other major universities, this array of courses forms a kind of grand quasiexperiment, the results of which hint at possible explanations for GPDs and suggest an array of future observations and experiments. 

In what follows, we provide a detailed explanation of our methods for quantifying student performance and measuring GPDs. We then describe patterns observed in the distribution of GPDs across the courses we study, discuss possible explanations for these patterns, and conclude with suggestions for future study and action.  

\section{\label{sec:level2}Performance signals, grade anomalies, and GPDs}

Grades provide students with their primary performance feedback: signals which affect their academic choices. Grades are also the only performance measures uniformly maintained and reported by nearly all colleges and universities. They are signals institutions use to mark course completion, determine academic standing, and award degrees and honors. To better understand these signals, we have compared grades students receive in each class to simple expectations they might hold. We begin by forming these expectations as they do: using grades received in other classes. Two important features emerge from this analysis. First, STEM lecture courses impose substantial grade penalties on all students. Second, these penalties are significantly larger for female than male students.

To examine these grade signals, we have assembled extensive student record data from the University of Michigan; including grades, GPAO (GPA in other classes at Michigan), high school GPA, standardized test scores, and demographic information. We gather these data for all students in 116 ‘large’ courses: those with average fall/winter enrollments more than 200 during the period from Fall 2008-Winter 2015. The total dataset contains 627,998 individual student/course pairs, while each course is represented by anywhere from 1,379 to 22,871 individual students. These courses come from a wide range of disciplines, including Science, Engineering, and Mathematics, the Social Sciences, and the Humanities.

To introduce the analysis conducted here, we compare student grades in one chosen course to their GPAO: a credit hour weighted average of grades they have received in all other classes. We calculate GPAO using all other grades received at Michigan up to the time when the course grade of interest is awarded. This allows us to include almost all students, even those just completing their first term in college. Figure 1 shows the GPAO-dependent structure of grades for male and female students in a representative STEM lecture course: in this case Physics 140 - first semester physics for scientists and engineers. 

Grades awarded to students in physics, compared to those they receive in other courses, are anomalously low. When students sign up for Physics 140, they should know that they will, on average, receive lower grades than they're used to. We might say that this course imposes a ``grade penalty'' on students. Figure 1 also reveals that this penalty is substantially larger for female students than male students - the course exhibits a large gendered performance difference. 

Could factors other than GPAO acount for this difference? Both regression and optimal matching on college of enrollment, ACT Math, ACT English, and high school GPA have minimal effects on inferred size of this gendered performance difference (Section IIIA). Neither does the difference lie simply in gendered GPAO differences. Female students in this class have slightly higher GPAOs than male students, but they receive substantially lower physics grades.

Figure 2 displays grading patterns which we observe for four additional courses likely to have analogs on other campuses: General Chemistry (Fig. 2a) , Calculus I (Fig. 2b), Introduction to Psychology (Fig. 2c), and First Year Writing (Fig. 2d). This set of examples gives some idea of the diversity in relations between grade and GPAO. Grades are all most strongly correlated with GPAO, but both grade penalties and the gendered performance difference can be either positive or negative. Again, observed gendered performance differences weaken only marginally using standard regression and matching techniques on LASSO-based covariates (Section IIIA).

\section{\label{sec:level3}Patterns of grade anomaly and gendered performance difference}

To gain a broader perspective on the student experience with large courses, we have explored relations between grades and GPAO across the full array of 116 large courses in our data set. To make these comparisons, we begin by defining simple measures of average grade anomaly and gendered performance difference. We choose these straightforward measures for two reasons. First, they are natural: they reflect ways in which students themselves might assess performance. Second, they generate measures of GPD which closely mirror those produced using much more complex analytic approaches, while remaining easy to replicate on other campuses.

Since the inception of grading more than a century ago, commentators have lamented the lack of consistency in grading practice, noting that average grades vary substantially by instructor, discipline, and institution \cite{Meyer1908-uf,finkelstein1913marking,Hills1968-fb,Goldman1974-hp,Schinske2014-fw}. This variability remains widespread, and has inevitable impacts on students' lives. We explore the impact of variable grading practice on the experience of students by measuring the average grade anomaly (AGA) of each course:
\begin{equation}
\textrm{Average Grade Anomaly} = \langle \textrm{Grade-GPAO} \rangle_{All}
\end{equation}

When this quantity is positive, students on average receive grades higher than they’re used to: a grade bonus. When it is negative, students on average receive grades lower than they’re used to, experiencing a grade penalty. We choose to describe grade anomalies as `penalties' and `bonuses' in an effort to reflect the way they are experienced by students. We also compare the AGAs for male and female students to form an initial measure of gendered performance difference (GPD):
\begin{equation}
\textrm{GPD}=\textrm{AGA}_f-\textrm{AGA}_m
\end{equation}

When this quantity is positive, the relative performance of female students is better. When it is negative, the relative performance of male students is better. Grade anomalies form collective measures of the feedback students receive from a course. Gendered performance differences compare the feedback received by male and female students in the same course.

Average grade anomalies and gendered performance differences for all of the 116 large U-M courses in our data set are shown in Figure 3. Each is labeled as belonging to Science and Engineering, the Social Sciences, or the Humanities. In general, STEM courses impose the largest grade penalties on students. These are often accompanied by significant gendered performance differences. Courses in the Social Sciences span a broad range of grade anomalies, and generally exhibit small gendered performance differences. Humanities courses typically award modest grade bonuses, and like Social Science courses, exhibit small gendered performance differences. Details for each course are contained in Table 1 in the supporting information \cite{supplement}.

Average grade anomalies are imposed by disciplinary differences in grading practice. Faculty members collectively decide on norms for the average grades awarded in these classes. Core STEM disciplines and some Social Sciences have chosen to continue awarding low average grades during an era of substantial grade inflation. As a result, students taking these courses are awarded anomalously low grades; they pay grade penalties imposed by variations in the norms of grading practice across our institutions. It is unsuprising to find that lecture courses in the core STEM disciplines often impose large grade penalties on students. At Michigan, these are as large as -0.62 and averaging -0.22 letter grades (typical standard error $<$ 0.01). The presence of these grade penalties is widely acknowledged within the STEM education community \cite{sabot1991grade}.

What is less well known is that most of these large STEM courses also exhibit significant gendered performance differences: male students experience smaller grade penalties than female students. Many of these differences are substantial, as large as -0.28 and averaging -0.11 letter grades. Even after optimal matching the average GPD across all STEM courses is still -0.10. While gendered performance differences are much less pronounced in the Humanities and Social Sciences, first and second semester Economics courses cluster with the STEM lectures. 

Focusing our attention on STEM courses (Fig. 4), interesting features emerge. We begin with first year introductory physics, chemistry, biology, and mathematics lecture courses. These courses (indicated by stars in the figure) are the gateways to a STEM degree, required for completing any of these majors. At Michigan, most of these courses do not impose fixed grade curves, but they do have norms of practice which maintain low and stable average grades. As a result, every one of them imposes significant grade penalties on its students, with AGAs averaging -0.41 letter grades, and ranging from -0.54 to -0.15.  

These gateway STEM lecture courses also exhibit substantial gendered performance differences; the mean is -0.18, ranging from -0.28 to -0.07. Female and male students receive different average grades in these courses, even when enrolled in the same college, and having earned the same GPAO, standardized test scores, and HS GPA. As a result, female students face grade penalties substantially larger than men, averaging -0.51 letter grades, and ranging from -0.63 to -0.24. These gendered performance differences correspond to roughly 10\% differences in course point totals: they are both statistically and materially significant. While GPDs have long been noted in specific fields (e.g. \cite{kost2009characterizing}), their ubiquity across introductory STEM lecture courses has not been clearly recognized.

Interestingly, the lab courses associated with these the same STEM subjects show a very different pattern. Their grade anomalies are small, averaging 0.12 letter grades (a small bonus), and ranging from -0.16 to 0.35. Higher average grades are given in these courses; they are subject to different grading norms of practice. More strikingly, their gendered performance differences are both small and various, with a mean of 0.02, ranging from -0.1 to 0.06.  While the subject matter in lectures and labs is closely related, something about these two types of courses leads to quite different outcomes.

\subsection{Other measures of aptitude do not explain gendered performance differences}

Grade anomaly calculations use only GPAO to account for aptitude in predicting performance. Indeed, additional factors have been shown elsewhere to have predictive power for grades in large enrollment Life Science classes (e.g. \cite{creech2012}), and non-zero gendered performance differences demonstrate that other factors affect performance. It is possible that other measures of aptitude also predict performance, perhaps in ways which 'explain away' observed gendered performance differences. 

To test for this, we have conducted an extensive LASSO (e.g. \cite{efron2004,Hastie2005-nj} analysis of grade prediction in all of our 116 large courses (See Table 1 and the Supplemental Material \cite{supplement}). In nearly every case, GPAO is the most important predictor of grade. High school GPA and standardized tests scores add little comparative predictive power, and while they narrow gender performance differences somewhat, they call far short of eliminating those seen in our large intro STEM courses. Adjustment of grade predictions for these covariates by either basic linear regression or a variety of matching techniques all yield similar results; somewhat reducing gendered performance differences but not eliminating them.

\subsection{Evaluative style and GPD}

Our large introductory STEM lecture courses differ in many ways. Some are offered in relatively small sections and taught in a studio style (e.g. Math 115 and 116), others are large traditional lecture classes with relatively little active engagement (e.g. Chem 130, Bio 172). They share some traits as well: student grades in every one are largely determined by timed examinations, many of which use multiple choice and short answer formats. Such exams typically make up 70\% of course grades, and generate almost all of the dispersion among student grades. Lab courses, by contrast, all meet in small sections, and none are primarily evaluated using timed examinations. We speculate that this difference in evaluative scheme plays a role in generating the striking gendered performance differences we see in gateway STEM lecture courses.

The other courses we observe help to explore this hypothesis: details of all, including links to their online course descriptions, are provided in Table 1. Performance in many of the additional STEM courses is also evaluated in timed, high-stakes examinations. A series of other first year courses in Computer Science (EECS 183, 280, 281), Biological Chemistry (BIOLCHEM 212), and Genetics (BIOLOGY 305) show grade anomalies and gendered performance differences very much like those of the core STEM lecture courses. ENGR 101 is a programming course for first year engineers. It provides a small grade bonus, but still exhibits gendered performance differences. 

The EARTH courses and ASTRO 106 are all 1-2 credit lecture courses intended as natural science distribution credit for non-STEM majors. Perhaps because they are elective courses, their instructors choose not to impose significant grade penalties on their students. While they offer grade bonuses, they still exhibit substantial gendered performance differences. These large courses are also evaluated primarily by with timed, multiple-choice, examinations.

Some other STEM courses are evaluated in very different ways. ENGR 100, 110, and TCHNCLCM 300, while labeled as lectures, are quite different in practice. ENGR 100 is a project-based introduction to engineering, ENGR 110 a one credit seminar introducing engineering education and careers, and TCHNCLCM 300 is a writing-based course often structured around group work. Biology 118 is a course on AIDS for non-scientists. None of these courses used timed examinations as important evaluative elements.

Courses taken beyond the first year by students engaged in STEM majors sometimes show different behaviors. This is likely due to selection among the students who enroll. One example is provided by the mathematics sequence of courses MATH 115/116/215/216. Female students differentially depart from from this sequence throughout: they make up 44\% of MATH 115 students and only 26\% of MATH 216 students. The gradual shift in gendered performance difference in these courses may be due to these differential selection effects. 

Additional evidence for suggesting the importance of evaluative scheme is provided by internal grades in the physics courses studied here. In these courses, some credit (25-30\%) is awarded for electronic response to in class questions and successful completion of online homework assignments. In both categories, female physics students modestly outperform males, though all students receive most of the available credit. Most of each student’s physics grade (70-75\%) is awarded for performance on three midterm exams and a final. Scores on these timed exams show gendered performance differences ranging from 3.5 – 7\% across these courses. Differences in exam performance are the source of the gendered performance differences, at least in these physics courses. 

The apparent impact of course evaluative scheme on performance might be related to the gendered performance differences observed in SAT/ACT college entrance exams \cite{rosser1989sat, Bridgeman1991-hd, Wainer1992-uq, Stricker1993-va, Mau2001-tm}. These tests are high pressure, timed exams similar in format to those used to evaluate our large introductory classes shown in Figure 4. 

A relation between gendered performance differences on standardized tests and in these classes is also suggested by the manner in which ACT scores explain the gendered performance differences we observe. In most classes, adding ACT scores to the model does little to close the gendered performance gap. In these cases whatever caused the gendered performance differences on the ACT does not affect college course performance in the same way. Only when the work done in classes is closely aligned with standardized tests do ACT scores explain the gendered performance differences observed in our classes. This effect is clearest in Math 105 (Precalculus) and Math 115 (Calculus I). According to our LASSO analysis (Supplemental Material), ACT MATH is the best predictor MATH 105 grade, and nearly equal in predictive power to GPAO for MATH 115. This is not true for any of the other STEM courses.

Gendered performance differences have been present on standardized tests in the US since their inception. This work shows that gendered performance differences also emerge in University STEM courses which use timed exams for evaluation, at levels which are not accounted for by the differences previously observed in standardized testing. 

\section{\label{sec:level4}Discussion}

The patterns of grade anomaly that we report emerge from differences in grading practice among the disciplines, and are consistent with those observed in previous studies. At the course level, grade anomalies provide signals to students. These signals guide self-assessment and may, along with other factors \cite{eccles1987gender}, shape selection of future courses and major field. 

That these grade anomalies are associated with gender, even after adjustment for a host of factors, compels us to understand why. The 1964 Civil Rights Act prohibits discrimination on the basis of sex. Even in the absence of different treatment, the disparate impact of a practice or environment can form a legal cause of action. These patterns of widespread gendered performance difference may compel us to change long-standing practices \cite{erman2015stereotype}.

One important next step is to explore whether the patterns of grade anomaly and gendered performance difference observed here at U-M are replicated on other campuses. In early 2014, we launched a Sloan Foundation funded learning and research analytics project aimed at exploring exactly this question. A follow-on paper \cite{Matz2016-gendered} will report on comparisons of grade anomalies and gendered performance differences in introductory courses across five major universities. These analyses are relatively simple to conduct. We encourage our colleagues at other institutions to examine student outcomes like this on their own campuses and share the results.

On average, all students pay grade penalties when taking large introductory STEM courses. It has been suggested that female students exhibit increased grade sensitivity in economics courses \cite{dynan1997underrepresentation,Rask2008-il,goldin2013notes} and the physical sciences in general \cite{Ost2010-qj}, so that grade penalties per se are more likely to drive them away from these courses and majors. The ubiquity of gendered performance differences in these courses significantly exacerbates this problem. The impact of grade penalty on persistence for male and female students in STEM disciplines in this data set will be explored in detail in a forthcoming paper.

The apparent relation between gendered performance difference and evaluative style leads us to hypothesize that stereotype threat \cite{Steele1995-os,steele1997threat,betz2013gender} may play an important role. If female students in these courses expect to confirm a negative stereotype about the performance of women on introductory STEM exams, they may expend some cognitive resources on this concern, reducing their performance at the $\sim$10\% level often imposed by stereotype threat \cite{Spencer1999-fx,Walton2009-is}. If so, actions aimed at reducing stereotype threat, ranging from values-affirmation to mindset interventions might be effective at eliminating these gendered performance differences \cite{walton2014new}. 

Efforts of this kind have sometimes shown promise \cite{walton2015two,miyake2010reducing}, and occasionally failed \cite{wilson2011redirect}. Given the very widespread nature of these gendered performance differences, supportive interventions need to be applied at scale, across a wide range of disciplines. Recent work suggests that such large scale interventions, delivered online, can be effective \cite{paunesku2015mind}. They should be pursued in the near term, as a way of ameliorating the effects of stereotype threat while these courses retain timed examinations as an evaluative tool. 

The focus here has been placed on performance differences associated with gender. It is important to recognize that performance gaps associated with other aspects of student identity can also be studied using the approaches described in this paper. These analyses, to be reported in detail in a subsequent paper, reveal performance differences both larger and smaller than the GPDs reported here. They may be associated with related causes.

While many steps might be taken to ameliorate these GPDs, we suggest that a more comprehensive shift in assessment style is needed \cite{duschl2016reconceptualizing}. The best approach may be a wholesale shift toward evaluation using more scientifically authentic activies and away from a dominant reliance on inauthentic timed examinations. 

The substantial grade anomalies observed for required first year STEM courses provide a barrier to entry for all students. Substantial gendered performance differences exacerbate these challenges for female students. If adoption of an inauthentic evaluative style has imposed these additional barriers, we should experiment with changes which might eliminate them. 

\section{Acknowledgements}

This work has been supported by the NSF WIDER grant DUE-1347697 for the REBUILD project, by NSF TUES grant DUE-1245127, and by the University of Michigan Provost's Learning Analytics Task Force through the Learning Analytics Fellows Program. We thank David Gerdes, August Evrard, Alexis Knaub, Kate Miller, Madeline Huberth, Eric Bell, Ben Hansen, and Rebecca Matz for suggestions and preliminary work which inspired this analysis. We also thank the U-M Registrar Paul Robinson, the Office of the Registrar, U-M CIO Laura Patterson, and all the staff at the U-M Information Technology Services division for both maintaining and supporting access to this remarkable data set. Finally, we acknowledge the important contributions of former U-M Provost Phil Hanlon and current U-M Provost Martha Pollack. Their strong advocacy of appropriate research using student record data has made learning analytics at Michigan possible. This research has been determined exempt from human subjects control under exemption \#1 of the 45 CFR 46.101.(b) by the U-M Institutional Research Board (HUM00079609). 

\onecolumngrid 

\begin{figure}[]
  \centering
    \includegraphics[width=1.0\textwidth]{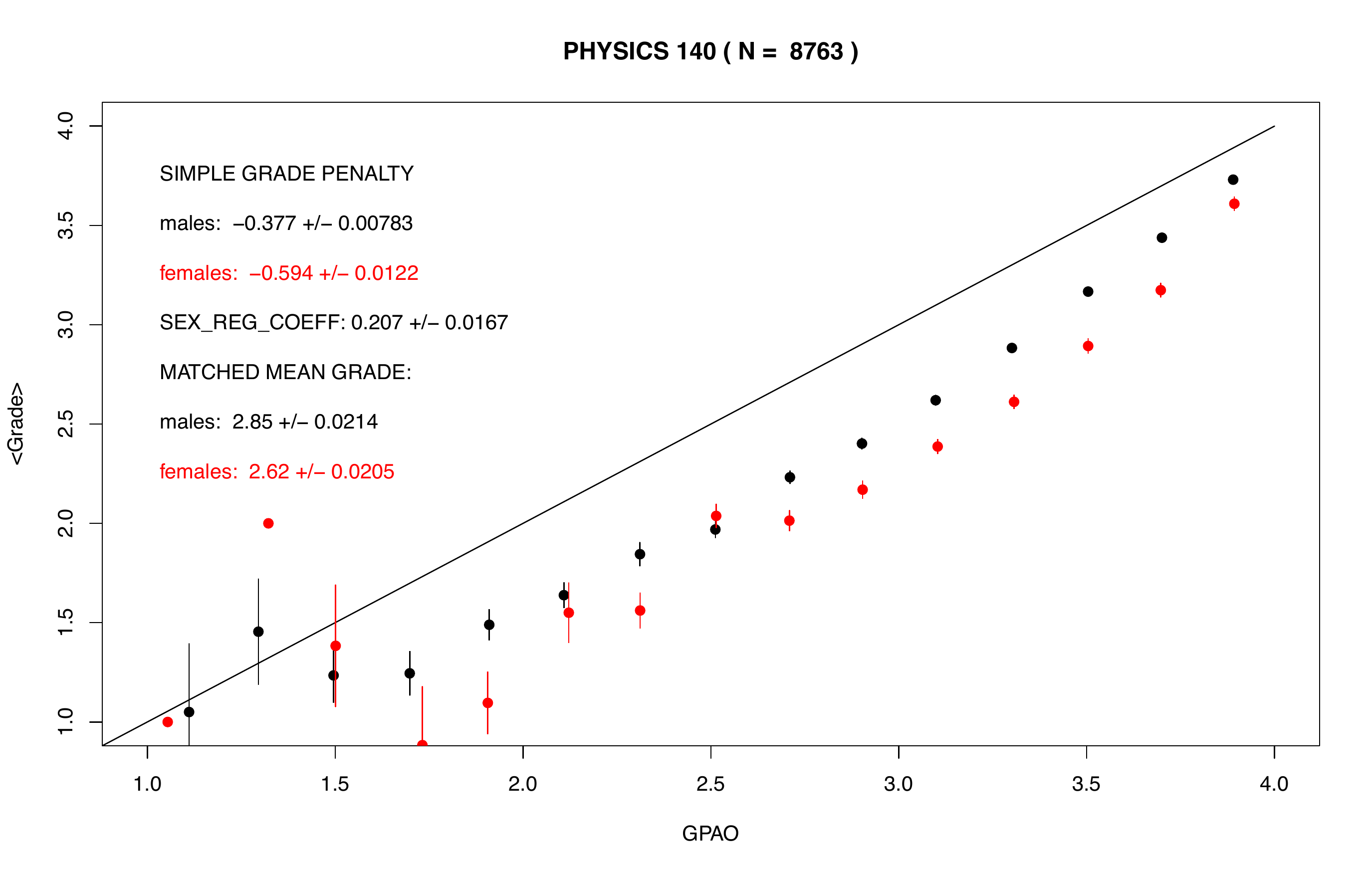}
  \caption{\bf{Grade Anomaly in Introductory Physics}. Male (black) and female (red) students are binned by GPA in other classes (GPAO), and the mean grade and standard error (bootstrap resampling, see Supporting Information) are plotted, along with a line of equality. The average grade anomaly is listed separately for male and female students, along with the results of regression and matching measures of GPD.} \label{grade_penalty.fig}
\end{figure}

\begin{figure}[]
  \centering
  \begin{tabular}{cc}
   \includegraphics[width=0.45\textwidth]{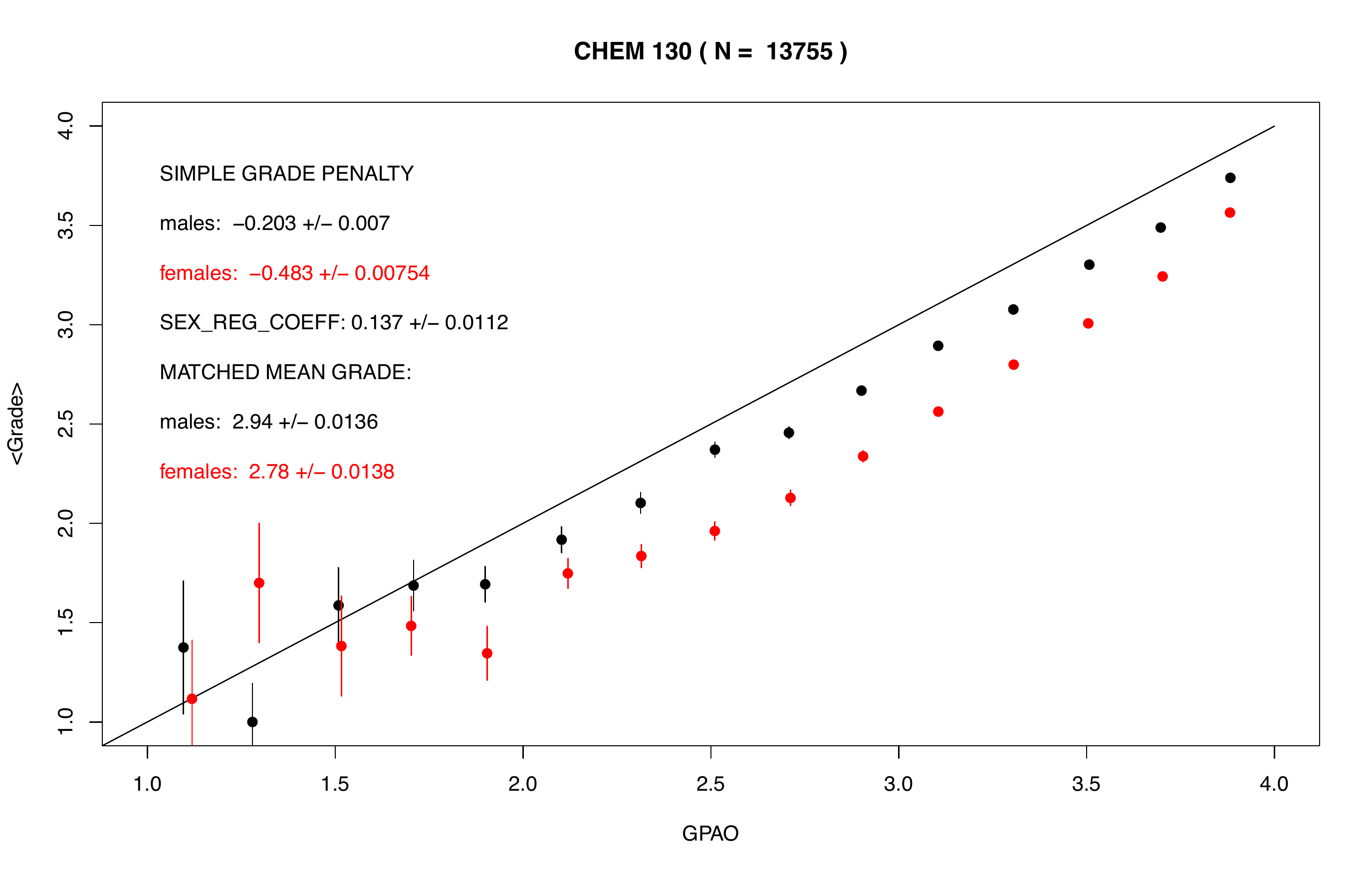}&
    \includegraphics[width=0.45\textwidth]{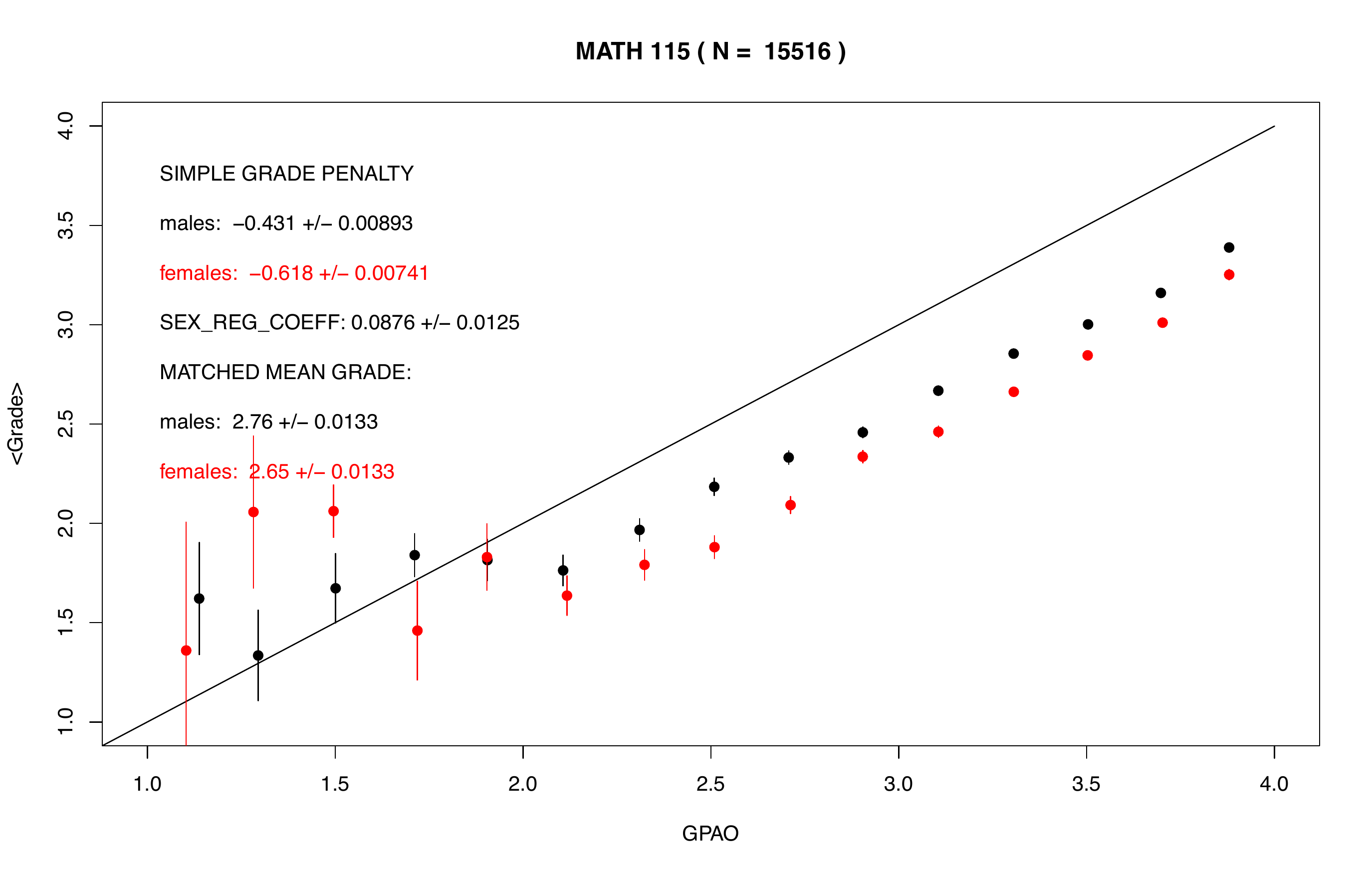}\\
    \includegraphics[width=0.45\textwidth]{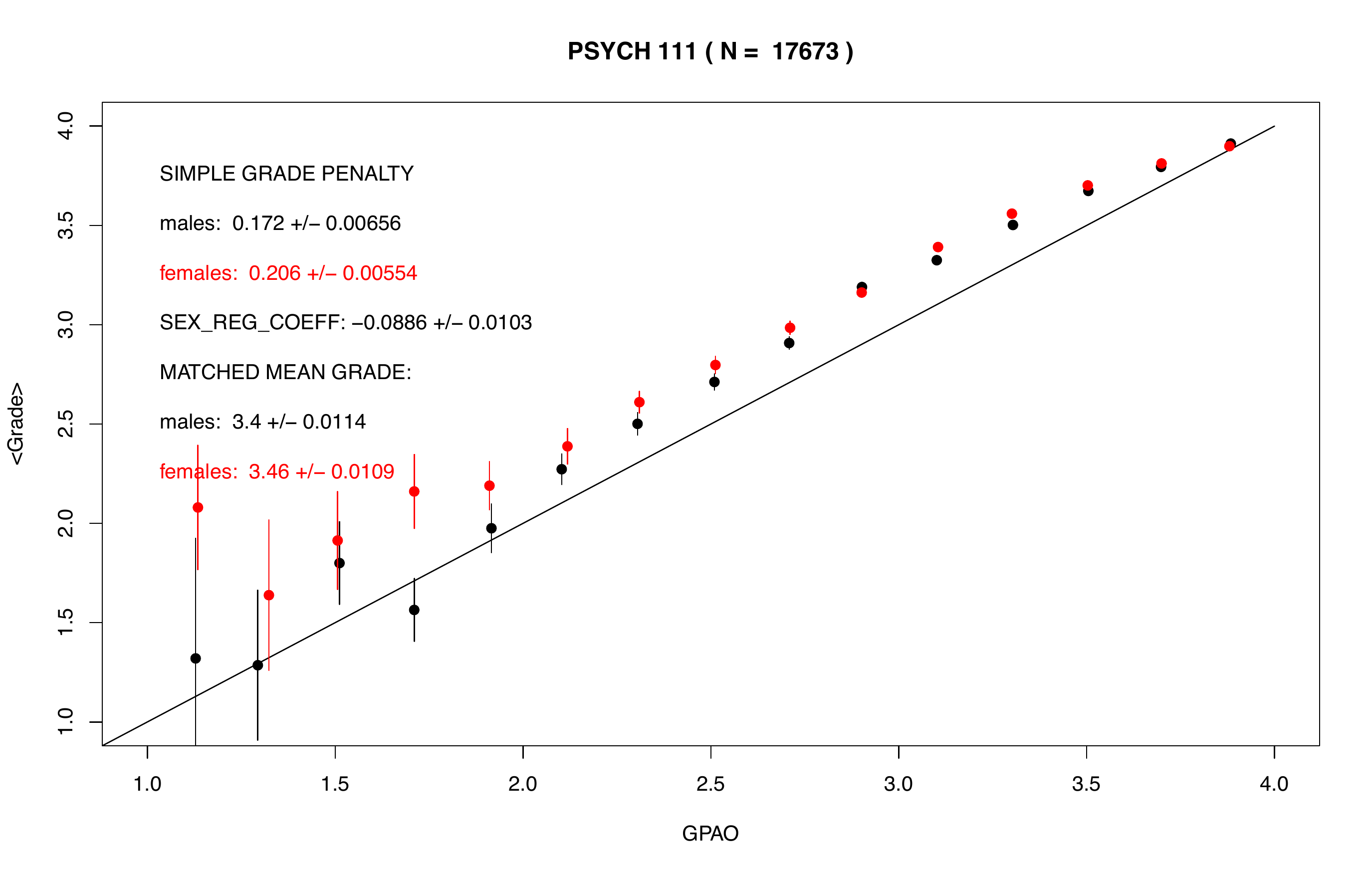}&
    \includegraphics[width=0.45\textwidth]{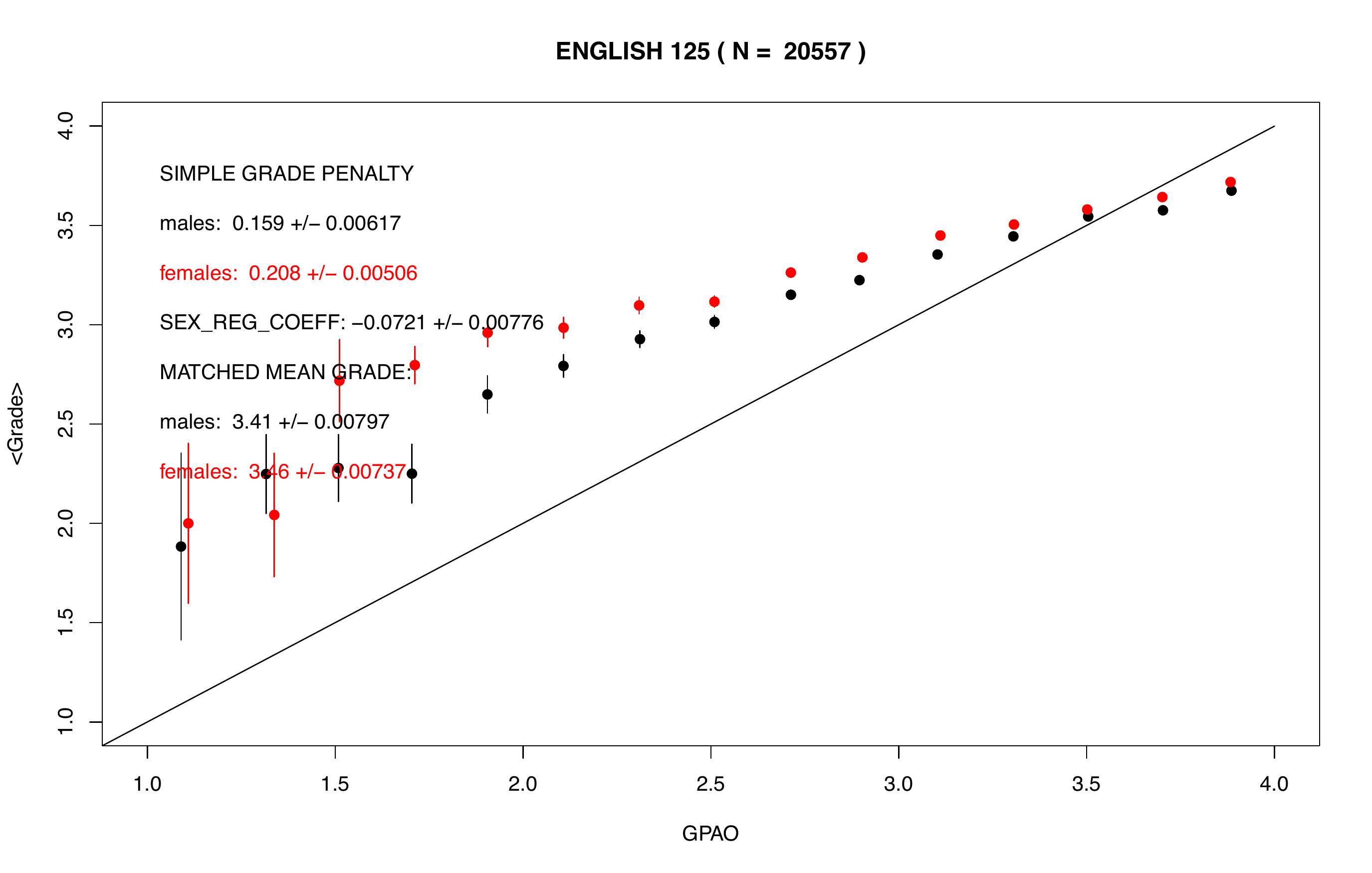}\\
  \end{tabular}
  \label{figure}\caption{\bf{Examples of Grade Anomalies in four introductory courses at U-M} General Chemistry (a), Calculus I (b), Introduction to Psychology (c), and First Year Writing (d).}\label{grade_penalty_ex.fig}
\end{figure}

\begin{figure}[h!]
  \centering
    \includegraphics[width=1.0\textwidth]{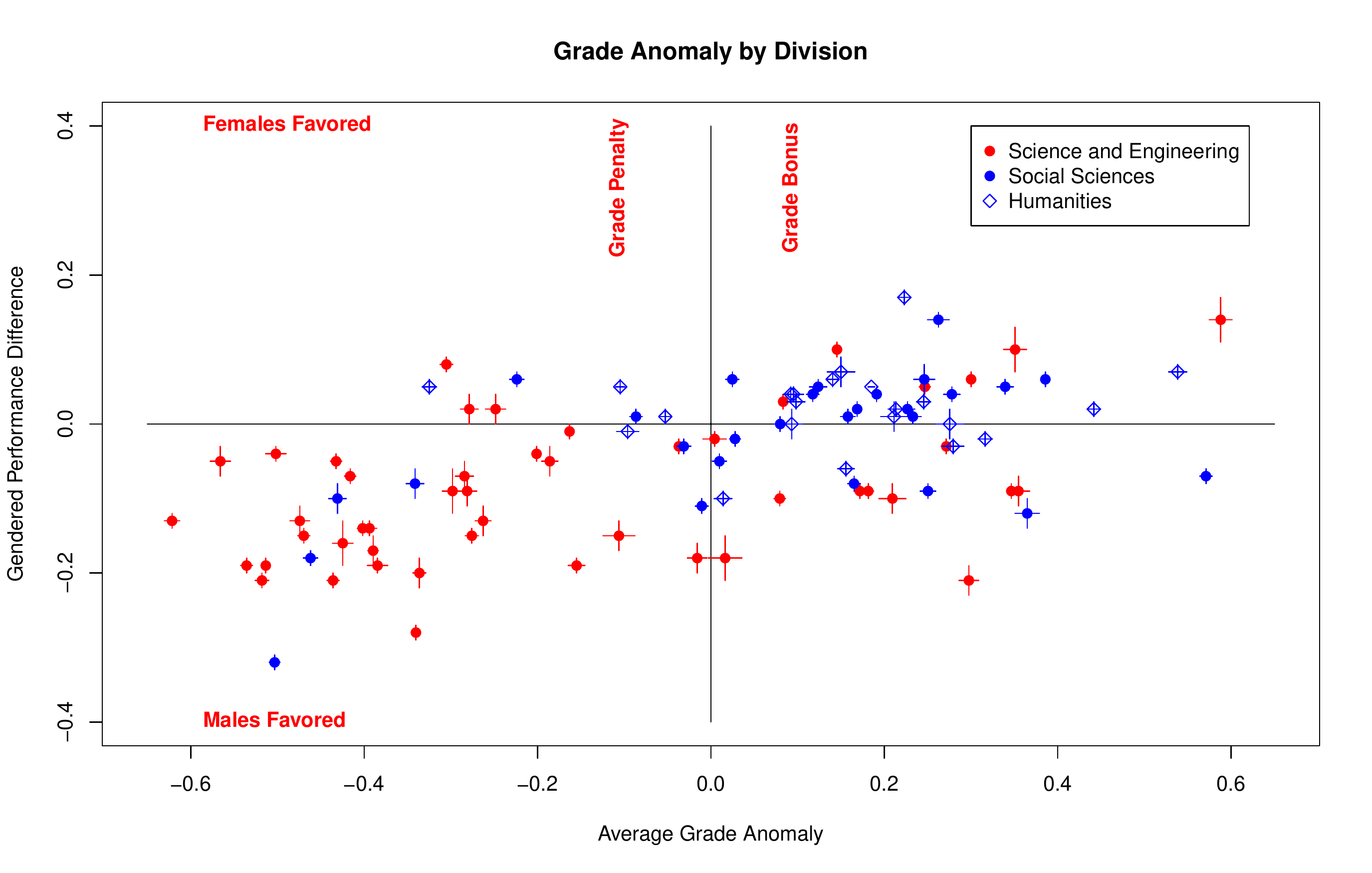}
  \caption{\bf{Grade Anomalies University-Wide}. Gendered performance difference is plotted against average grade anomaly for each of 116 courses. Different symbols divide courses into three divisions: science and engineering (red filled circles), social sciences (blue filled circles), and humanities (blue diamonds). Errors bars are the SE on the mean as determined by bootstrap resampling.} 
  \label{all_divisions.fig}
\end{figure}

\begin{figure}[h!]
  \centering
    \includegraphics[width=1.0\textwidth]{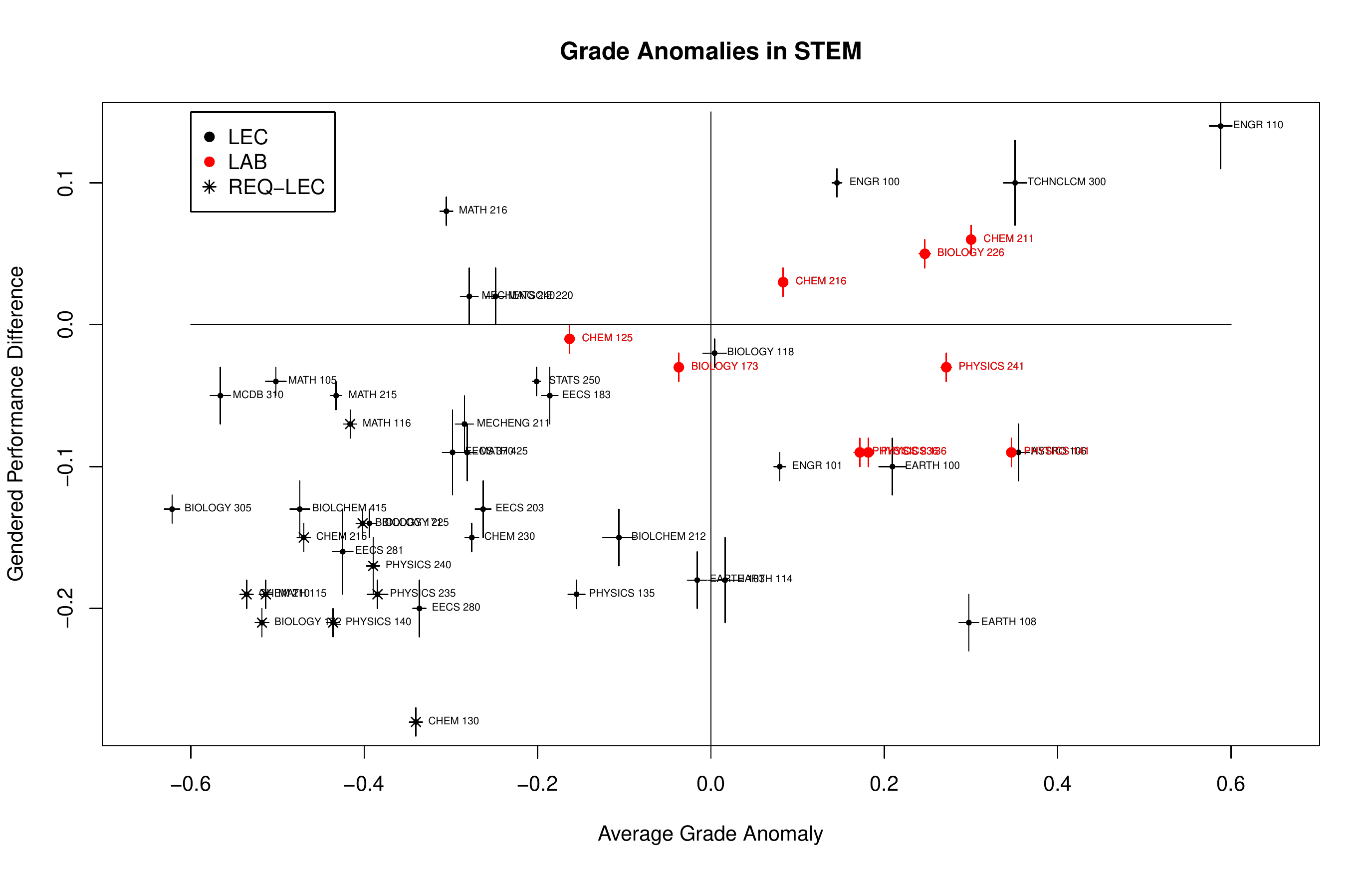}
  \caption{\bf{Grading Trends in Introductory STEM courses.} Color differentiates lectures and labs; stars represent core required STEM lecture courses.}\label{stem.fig}
\end{figure}

\twocolumngrid 

\bibliography{bibliography}

\end{document}